\documentclass[11pt]{amsart}
\usepackage{amssymb, latexsym, amsmath}
\usepackage{graphicx}
\usepackage{color}

\newcommand{\R}{\mathbb R}

\begin{document}
\title[Quasi-local mass]{Quasi-local mass at axially symmetric null infinity}
\author{Po-Ning Chen, Mu-Tao Wang, Ye-Kai Wang, and Shing-Tung Yau}
\date{\today}

\begin{abstract}
We give a brief review of the definition of the Wang-Yau quasilocal mass and discuss the evaluation of which on surfaces of unit size at null infinity of an axi-symmetric spacetime in Bondi-van der Burg-Metzner coordinates. \end{abstract}

\thanks{P.-N. Chen is supported by NSF grant DMS-1308164 and Simons Foundation collaboration grant \#584785, Y.-K. Wang is supported by MOST Taiwan grant 105-2115-M-006-016-MY2, 107-2115-M-006-001-MY2, and S.-T. Yau is supported by NSF grants  PHY-0714648 and DMS-1308244. This material is based upon work supported by the National Science Foundation under Grants No. DMS-1405152 and No. DMS-1810856 (Mu-Tao Wang).  Part of this work was carried out when the authors were visiting the National Center of Theoretical Sciences at National Taiwan University in Taipei, Taiwan. }

\address{Po-Ning Chen\\Department of Mathematics\\University of California, Riverside, USA}
\email{poningc@ucr.edu}

\address{Mu-Tao Wang\\Department of Mathematics\\
Columbia University\\ New York\\ NY 10027\\ USA}
\email{mtwang@math.columbia.edu}

\address{Ye-Kai Wang\\Department of Mathematics\\National Cheng Kung University, Taiwan}
\email{ykwang@mail.ncku.edu.tw}

\address{Shing-Tung Yau\\Department of Mathematics\\Harvard University, USA}
\email{yau@math.harvard.edu}

\maketitle
\date{\usdate{\today}}
%\date{XXX XX, 20XX}

\section{Introduction}
 As is well known, it is not possible to find mass density of gravity
in general relativity. The mass density would have to consist of first
derivatives of the metric tensor which are zero in suitable chosen
coordinate at a point. But we still desire to measure the total mass in a spacelike region
bounded by a closed surface. The mass due to gravity should be computable from the intrinsic
and the extrinsic geometry of the surface. It has been considered to be one of the important questions to find the right definition.
Penrose gave a talk  \cite{Penrose1} at the Institute
for Advances Study in 1979 and listed it as the first one in his list of major open
problems. The quantity is called quasilocal mass. 

Many people including Penrose \cite{Penrose2}, Hawking \cite{Hawking}, Brown-York \cite{BY2}, Hawking-Horowitz \cite{HH}, Bartnik \cite{Bartnik2}
and others worked on this problem and various definitions have been
given. Several important contributions have also been made by Bartnik \cite{Bartnik1}, Shi-Tam \cite{Shi-Tam}, and Liu-Yau \cite{Liu-Yau1, Liu-Yau2}.
This article discusses the Wang-Yau quasilocal mass definition discovered in 2009 \cite{Wang-Yau1, Wang-Yau2}.

\section{The definition of Wang-Yau quasilocal mass}
Quasilocal mass is attached to a 2-dimensional spacetime surface which is a topological 2-sphere (the boundary of a spacelike region).
But with different intrinsic geometry and extrinsic geometry, we expect to read off the effect of gravitation in the spacetime vicinity of the surface. 
Suppose the surface is spacelike, i.e. the induced metric $\sigma$ is Riemannian.  An essential part of the extrinsic geometry is measured by the mean curvature vector field $\bf{H}$, which is a normal vector field of the surface such that the null expansion along any null normal direction $\ell$ is given by the paring of $\bf{H}$ and $\ell$.

To evaluate the quasilocal mass of a 2-surface $\Sigma$ with the physical data $(\sigma, \bf{H})$, one solves the {\it optimal isometric embedding system} (see \eqref{oiee} in the next paragraph), which gives an embedding of $\Sigma$ into the Minkowski spacetime with the image
surface $\Sigma_0$ that has the same 
induced metric $\sigma$ as $\Sigma$. We then compare the extrinsic geometries of $\Sigma$ and $\Sigma_0$ and evaluate the quasilocal mass from $\sigma, \bf{H}$, and $\bf{H_0}$.

 The physical surface $\Sigma$ with physical data $(\sigma, \bf{H})$ gives $(\sigma, |\bf{H}|, \alpha_{\bf {H}})$, where 
 $|\bf{H}|$ is the norm of the mean curvature vector field and  $\alpha_{\bf {H}}$ is the connection one-form determined by the mean curvature gauge. As long as the mean curvature vector field ${\bf H}$ is spacelike, $|{\bf H}|$ is positive and $\alpha_{\bf H}$ is well-defined. 
 Given an isometric embedding $X:\Sigma\rightarrow \R^{3,1}$ of $\sigma$, let $\Sigma_0 = X(\Sigma)$ be the image and $(\sigma, |\bf{H}_0|, \alpha_{\bf {H}_0})$ be the corresponding data of $\Sigma_0$. 
 Let $T$ be a future timelike unit Killing field of $\R^{3,1}$ and define $\tau=-\langle X, T\rangle$ as a function defined on the surface $\Sigma$ where $\langle \cdot, \cdot \rangle$ is the Minkowski metric on $\R^{3,1}$. 

 The optimal isometric embedding system and the quasilocal mass can be expressed in terms of  a function $\rho$ and a 1-form $j_a$ on $\Sigma$ given by 
  \[ \begin{split}\rho &= \frac{\sqrt{|{\bf H_0}|^2 +\frac{(\Delta \tau)^2}{1+ |\nabla \tau|^2}} - \sqrt{|{\bf H}|^2 +\frac{(\Delta \tau)^2}{1+ |\nabla \tau|^2}} }{ \sqrt{1+ |\nabla \tau|^2}}\\
 j_a&=\rho {\nabla_a \tau }- \nabla_a \left( \sinh^{-1} (\frac{\rho\Delta \tau }{|\bf H_0||{\bf H}|})\right)-(\alpha_{\bf H_0})_a + (\alpha_{{\bf H}})_a, \end{split}\]  where $\nabla_a$ is the covariant derivative with respect to the metric $\sigma$, $|\nabla \tau|^2=\nabla^a \tau\nabla_a \tau$ and $\Delta \tau=\nabla^a\nabla_a \tau$. 
 
 The optimal isometric embedding system seeks for a solution $(X, T)$ that satisfies  
\begin{equation}\label{oiee}\begin{cases}
\langle dX, dX\rangle&=\sigma\\
\nabla^a j_a&=0.
\end{cases}\end{equation} We note that the first equation is the isometric embedding equation into the Minkowski spacetime $\R^{3,1}$.  The quasilocal mass is then defined to be \begin{equation}\label{qlm} E(\Sigma, X, T)=\frac{1}{8\pi}\int_\Sigma \rho.\end{equation}

Several remarks are in order:

(1) $\Sigma_0$ is the ``unique" surface in the Minkowski spacetime that best matches the physical surface $\Sigma$. 
If $\Sigma$ happens to be a surface in the Minkowski spacetime, the above procedure identifies $\Sigma_0=\Sigma$ up to a global isometry.

(2) A prototype form of the quasilocal mass which corresponds to the special case $\tau=0$ in \eqref{qlm} (due to Brown-York \cite{BY2}, Liu-Yau \cite{Liu-Yau1}, Booth-Mann \cite{Booth-Mann}, Kijowski \cite{KI}) is 
\[\frac{1}{8\pi}\int_\Sigma \left(|{\bf H_0}|-|\bf H|\right).\]  The positivity is proved by Shi-Tam \cite{Shi-Tam} and Liu-Yau \cite{Liu-Yau2}. However, for a surface in the Minkowski spacetime, the above expression may not be zero \cite{OST}. 

 The optimal isometric embedding system gives the necessary correction, so that the Wang-Yau definition is positive in general and zero for surfaces in the Minkowski spacetime \cite{Wang-Yau1, Wang-Yau2}. 

 The derivation of the Wang-Yau definition \eqref{qlm} relies on both physical theory and mathematical theory.  From the Hamilton-Jacobi analysis of the Einstein-Hilbert action (Brown-York \cite{BY2}, Horowitz-Hawking \cite{HH}), a surface Hamiltonian $\mathfrak{H}(\Sigma)$ is obtained and the quasilocal energy should be $\mathfrak{H}(\Sigma)-\mathfrak{H}(\Sigma_0)$ for a reference surface $\Sigma_0$ in the reference Minkowski spacetime. The precise definitions of the surface Hamiltonians still depend on the choices of a normal gauge and a timelike vector field along the surface $\Sigma$ as an observer.
 
 On the other hand, the mathematical theory of isometric embeddings (Nirenberg \cite{Nirenberg}, Pogorelov \cite{Pogorelov}) is used to find the reference surface $\Sigma_0$ and a variational approach leads to a canonical gauge that anchors the choices of gauge of surface Hamiltonian.

In general, the optimal isometric embedding system is difficult to solve.   Suppose $(X, T)$ is a solution and suppose the corresponding $\rho$ is positive, then $E(\Sigma, X, T)$ is a local minimum \cite{Chen-Wang-Yau2} and the nearby
optimal isometric embedding system is solvable by an inverse function theorem argument. In a perturbative configuration, when a family of surfaces limit to a surface in the Minkowski spacetime, then the optimal isometric embedding system is solvable, again subject to the positivity of the limiting mass. This applies to the case of large sphere limits \cite{Chen-Wang-Yau1} and  small sphere limits \cite{Chen-Wang-Yau5}.

\section{Quasilocal mass at null infinity}

\subsection{Large sphere limit at null infinity}

Consider an isolated system surrounding a source. In terms of the Bondi-Sachs coordinate system $(u, r, x^2=\theta, x^3=\phi)$ \cite{BVM, S, Trautman1, Trautman2}, near future null infinity $\mathfrak{I}^+$ the spacetime
metric takes the form:
\[-V du^2-2U du dr+\sigma_{ab}(dx^a +W^a du)(dx^b+W^b du), a, b=2, 3\] such that each $V$, $U$, $W^a$ admit expansions in terms of integral powers of $r$ with \[\begin{split} V&=1-\frac{2m(u, \theta, \phi)}{r}+ O(r^{-2})\\
U&=1+ O(r^{-2})\\
\sigma_{ab} dx^a dx^b&=r^2(d\theta^2+\sin^2\theta d\phi^2)+2r (c d\theta^2-c\sin^2 \theta d\phi^2+2\underline{c} \sin \theta d\theta d\phi)+ O(1)\\
W^a&= O(r^{-2}) \end{split}\] The Bondi-Trautman mass is \begin{equation}\label{BT_mass} M_{BT}(u)=\frac{1}{4\pi}\int_{S^2} m(u, \theta, \phi),\end{equation} where $m(u, \theta, \phi)$ is
the mass aspect function defined at $\mathfrak{I}^+$.  As a result of the vacuum Einstein equation, the mass loss formula states 

\begin{equation}\label{mass_loss} \frac{d}{du} M_{BT}(u)=-\frac{1}{4\pi} \int_{S^2} \left( c_u^2+(\underline{c}_u)^2\right) \leq 0.\end{equation} This important formula represents the first theoretical verification of gravitational wave/radiation in the nonlinear setting.

 In \cite{Chen-Wang-Yau1}, we evaluate the large sphere limit of quasilocal mass which recovers the Bondi-Trautman mass. At a retarded time $u=u_0$, we consider the family of large spheres $\Sigma_r$ parametrized by an affine parameter $r$. The positivity of the Bondi-Trautman mass guarantees the unique solvability of the optimal isometric embedding system \eqref{oiee} with a solution $(X_r, T_r)$. Suppose $X_r$ and  $T_r$ admit 
expansions:
\[ T_r=T^{(0)}+
\sum_{k=1}^\infty T^{(-k)}r^{-k}\]
\[X_r=rX^{(1)}+X^{(0)}+\sum_{k=1}^\infty X^{(-k)}r^{-k},\] then 
$T^{(0)}$ is shown to be proportional to the Bondi-Sachs energy-momentum and $T^{(0)}$ being future timelike makes $T^{(-k)}$ and $X^{(-k+1)}$ solvable inductively for $k=1, 2\cdots$.

\subsection{Unit sphere limits at null infinity}

Both the Bondi-Trautman mass and the mass loss formula are global statements about 
$\mathfrak{I}^+$, i.e. they require the knowledge of all directions of $(\theta, \phi)$. 
The limit of quasilocal mass introduced in the following provides a quasilocal quantity along a single direction $(\theta, \phi)$ at $\mathfrak{I}^+$.

 Consider a null geodesic $\gamma (d)$ that approaches $\mathfrak{I}^+$. Around each point on $\gamma(d)$, consider a geodesic 2-sphere $\Sigma_d$ of unit radius.  The geometry of $\Sigma_d$ approaches the geometry of a standard unit round sphere of $\R^3$. 

 In the limit $d \rightarrow \infty$, we obtain two quantities.
 The first one is $\lim_{d\rightarrow \infty} E(\Sigma_d)$ which is of the order of $\frac{1}{d^2}$ with $E(\Sigma_d)\geq 0$. The second one is obtained by exploiting the vanishing of the $\frac{1}{d}$ term and  appears as a loop integral on the limiting surface that is of the order of $\frac{1}{d}$.

Several cases have been computed:

(1)  Linear gravitational perturbation of the Schwarzschild black hole \`a la Chandrasekhar. The linearized vacuum Einstein equation is solved by separation of variables
and solutions of linearized waves are obtained. The optimal isometric embedding system can be solved and the quasilocal mass can be evaluated by solving
\begin{equation}\label{ode} \begin{split}\Delta (\Delta+2)\tau&=\text{physical  data},\\
(\Delta+2)N&=\text{physical  data},\end{split}\end{equation} where $\tau$ and $N$ are functions on the standard 2-sphere and $\Delta$ is the Laplace operator. All distinctive features of the waves such as frequency and mode parameters are recovered.

(2) The Vaidya spacetime \footnote{The mass aspect function used here differs to that of \cite{Vaidya} by a factor of $2$.}

\[-(1-\frac{2m(u)}{r})du^2-2dudr+r^2 (d\theta^2+\sin^2\theta d\phi^2)\]

 The quasilocal mass of a unit sphere approaching null infinity is
\begin{equation}\label{qlm_Vaidya} E(\Sigma_d)=-\frac{1}{8 \pi d^2} \int_{S^2} (\partial_u m) \sin^2(\hat{\theta})+l.o.t. \geq 0\end{equation} The positivity of quasilocal mass corresponds to the mass loss formula in the Vaidya case. 

 One may expect that the limit of quasilocal mass in the direction of $(\theta_0, \phi_0)$ will recover the value of the mass aspect function $m(u, \theta_0, \phi_0)$. But notice that 
the mass aspect function is not pointwise positive, only the integrated Bondi-Trautman mass is positive by Schoen-Yau \cite{SY} and Horowitz-Perry \cite{HP}. In the Vaidya case, the mass aspect is recovered from the loop integral.

\section{Unit sphere limit in BVM coordinate near null infinity}

The Vaidya spacetime is a model for the Einstein-null dust system. In particular, the condition $-\partial_u m\geq 0$ in \eqref{qlm_Vaidya} can be interpreted as local energy condition for matters. In order to study the contribution of quasilocal mass at the  purely gravitational level, we study the null infinity of a vacuum spacetime. 
We discuss the case of an axi-symmetric spacetime in the Bondi- van der Burg-Metzner coordinates near null infinity. 
 The leading terms of the spacetime metric are of the form:
\[-(1-\frac{2\bf{M}}{r})du^2-2 dudr-2{\bf U} dud\theta+(r^2+2r{\bf C})d\theta^2+(r^2-2r{\bf C})\sin^2\theta d\phi^2,\]
where ${\bf M}, {\bf U}$, ${\bf C}$ are functions of $u$ and  $\theta$.

The $\frac{1}{d^2}$ term of the quasilocal mass of a unit sphere approaching null infinity is (up to a constant factor)

\begin{equation}
\label{bvm}\int_{B^3} [(\partial_u {\bf C})^2+\det (h_0^{(-1)}-h^{(-1)})]+\frac{1}{4}\int_{S^2}[ (tr_\Sigma k^{(-1)})^2-\tau^{(-1)}\tilde{\Delta}(\tilde{\Delta}+2)\tau^{(-1)}],\end{equation} in which $h^{(-1)}$ and $k^{(-1)}$ depend on the physical data and $h_0^{(-1)}$ and $\tau^{(-1)}$ depend on the solution of the optimal isometric embedding system. A priori, the expression may depend on all ${\bf M}, {\bf U}$, and ${\bf C}$. However all occurrences of ${\bf M}$ and ${\bf U}$ are cancelled and the final answer only involves the function ${\bf C}$ and is completely independent of the mass aspect function ${\bf M}$. The expression \eqref{bvm} is manifestly positive by Wang-Yau's theorem.

\subsection{Final remark}

 Given a spacetime surface $(\Sigma, \sigma, \bf{H})$, we find the reference surface $(\Sigma_0, \sigma, \bf{H}_0)$ through the optimal isometric embedding equation and evaluate the quasilocal mass. This is a nonlinear and coordinate independent theory. The procedure is canonical and is accompanied by a uniqueness statement.  In particular, the definition does not involve any ad hoc referencing or normalization. The calculation of the quasilocal mass does not assume any a priori knowledge of null infinity. 

 The positivity of the unit sphere limit  should correspond to a ``quasilocal" mass loss formula, or ``quasilocal gravitational radiation" at null infinity. The evaluation in the full generality of Bondi-Sachs coordinates (without assuming axi-symmetry) will appear in an upcoming paper.

\end{document}